\documentclass[cits]{PoS}

\newcommand{\rgb}{RGB\,J0152$+$017}

\newcommand{\spectralindex}{$\Gamma=2.95$\-$\pm0.36_{\mathrm{stat}}$\-$\pm 0.20_{\mathrm{syst}}$}

\title{Discovery of VHE $\gamma$-rays from \rgb}

\ShortTitle{Discovery of VHE $\gamma$-rays from \rgb}

\author{\speaker{J.-P.~Lenain},$^a$ D.~Nedbal,$^b$ M.~Raue,$^c$ S.~Kaufmann,$^d$ L.~G\'erard,$^e$ M.~Hauser$^d$ and B.~Giebels$^f$ for the H.E.S.S. collaboration\\
  \llap{$^a$} Laboratoire Univers et Th\'eories (LUTH), Observatoire de Paris, CNRS, Universit\'e Paris Diderot, Meudon, France\\
  \llap{$^b$} Institute of Particle and Nuclear Physics, Charles University, Prague, Czech Republic\\
  \llap{$^c$} Max-Planck-Institut f\"ur Kernphysik, Heidelberg, Germany\\
  \llap{$^d$} Landessternwarte, Universit\"at Heidelberg, Heidelberg, Germany\\
  \llap{$^e$} Astroparticule et Cosmologie (APC), CNRS, Universite Paris 7 Denis Diderot, Paris, France\\
  \llap{$^f$} Laboratoire Leprince-Ringuet, Ecole Polytechnique, CNRS/IN2P3, Palaiseau, France\\
  E-mail: \email{jean-philippe.lenain@obspm.fr}, \email{dalibor.nedbal@mpi-hd.mpg.de}}

\abstract{
The BL\,Lac object \rgb\ ($z=0.080$) was predicted to be a very high-energy (VHE; $> 100$\,GeV) $\gamma$-ray source, due to its high X-ray and radio fluxes.

We report recent observations of this source made in late October and November 2007 with the H.E.S.S. array consisting of four imaging atmospheric \v{C}erenkov telescopes. Contemporaneous observations were made in X-rays with the {\textit{Swift\/}} and {\textit{RXTE\/}} satellites, in the optical band with the ATOM telescope, and in the radio band with the Nan\c{c}ay Radio Telescope.

As a result, \rgb\ is discovered as a source of VHE $\gamma$-rays by H.E.S.S. A signal of 173 $\gamma$-ray photons corresponding to a statistical significance of 6.6\,$\sigma$ was found in the data. The energy spectrum of the source can be described by a powerlaw with a spectral index of \spectralindex. The integral flux above 300\,GeV corresponds to $\sim$2\% of the flux of the Crab nebula. The source spectral energy distribution (SED) can be described using a two-component (extended jet and blob in jet) non-thermal synchrotron self-Compton (SSC) leptonic model, plus a thermal host galaxy component. The parameters that are found are very close to those found for TeV blazars in similar SSC studies.

The location of its synchrotron peak, as derived from the SED in {\textit{Swift\/}} data, allows clear classification as a high-frequency-peaked BL\,Lac (HBL).
}

\FullConference{Workshop on Blazar Variability across the Electromagnetic Spectrum\\
		 April 22-25 2008\\
		 Palaiseau, France}

\begin{document}

\section{The H.E.S.S. system}

\begin{figure}[h!]
  \centering
  \includegraphics[width=0.6\textwidth]{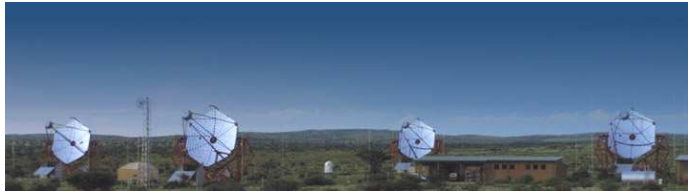}
  \caption{The H.E.S.S. \v{C}erenkov array located in Namibia.}
  \label{fig1}
\end{figure}

The H.E.S.S. (High Energy Stereoscopic System, see Fig.~\ref{fig1}) collaboration operates an array of four 13\,m imaging atmospheric \v{C}erenkov telescopes. The cameras measure $\gamma$-rays above a threshold of $\sim$100\,GeV up to several 10\,TeV by imaging the \v{C}erenkov light induced by an air shower developing when a VHE photon or particle enters the atmosphere.

\section{H.E.S.S. observations and results}

\begin{figure}[h!]
  \centering
  \includegraphics[width=0.6\textwidth]{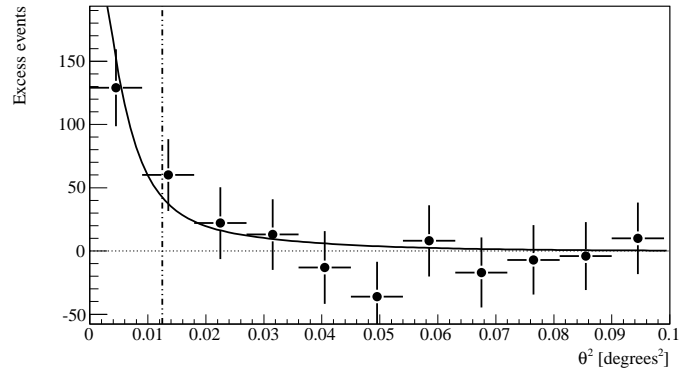}
  \caption{Angular distribution of excess events.}
  \label{fig2}
\end{figure}

The observations of \rgb\ by H.E.S.S. were performed between October 30 and November 14, 2007 \cite{2007ATel.1295....1N}. After applying quality cuts on data affected by poor weather conditions and hardware problems, the total live-time amounts to 14.7\,h (see \cite{2008A&A...481L.103A} for more details), with a mean zenith angle of the observations of 26.9$^\circ$. Using {\em standard cuts}, an excess of 173\,$\gamma$-ray events is found in the data, leading to a significance of the detection of 6.6$\sigma$ (see Fig.~\ref{fig2}).

\begin{figure}[h!]
  \centering
  \includegraphics[width=0.6\textwidth]{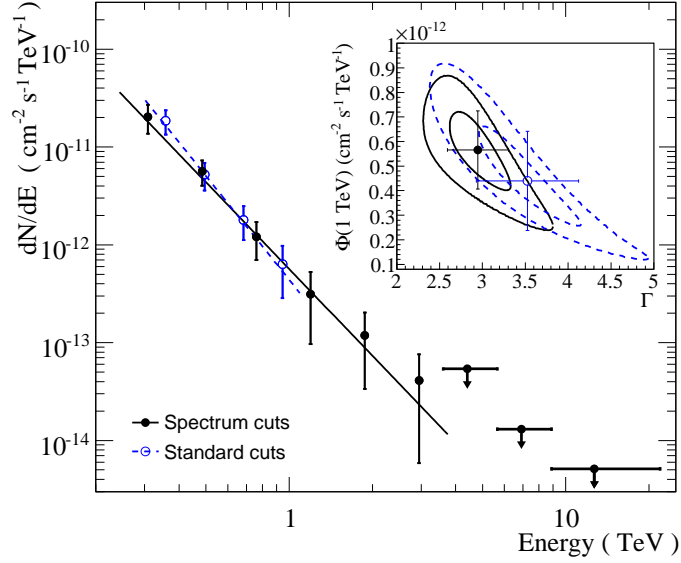}
  \caption{Differential spectrum of \rgb\ in the VHE range. The inlay shows 1 and 2$\sigma$ significance contours of the fit parameters.}
  \label{fig3}
\end{figure}

Figure~\ref{fig3} shows the time-averaged VHE differential spectrum, using both {\em standard cuts} above 300\,GeV and {\em spectrum cuts} (less restrictive, see \cite{2006A&A...457..899A}) between 240\,GeV and 3.8\,TeV. Using the latter cuts, the spectrum is described well ($\chi^2/\mathrm{d.o.f.}=2.16/4$) by a powerlaw d$N$/d$E = \Phi_0(E/$1\,TeV$)^{-\Gamma}$ with a photon index $\Gamma = 2.95$ $\pm 0.36_\mathrm{stat}$ $\pm 0.20_\mathrm{syst}$ and $\Phi_0 = (5.7$ $\pm 1.6_\mathrm{stat}$ $\pm 1.1_\mathrm{syst})$ $\times 10^{-13}$\,cm$^{-2}$\,s$^{-1}$\,TeV$^{-1}$.

All the VHE results here were cross-checked with independent analysis procedures and calibration chain, giving consistent results.

\begin{figure}[h!]
  \centering
  \includegraphics[width=0.6\textwidth]{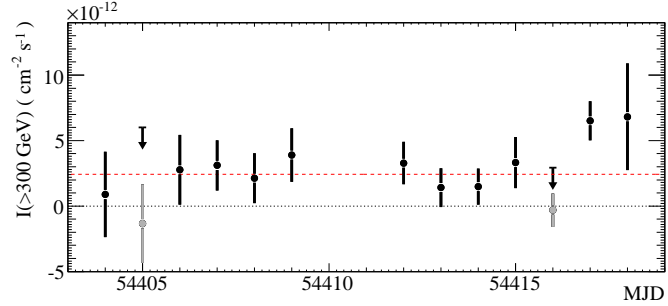}
  \caption{Mean nightly integral flux from \rgb\ above 300\,GeV. The red line shows the fit of a constant using all nights. Upper limits at 99\% confidence level are calculated when no signal is found (grey points).}
  \label{fig4}
\end{figure}

The integral flux above 300\,GeV is $I = (2.70 \pm 0.51_\mathrm{stat} \pm 0.54_\mathrm{syst}) \times 10^{-12}$\,cm$^{-2}$\,s$^{-1}$, corresponding to $\sim$2\% of the flux of the Crab nebula above the same threshold, as determined by \cite{2006A&A...457..899A}. The nightly evolution of the VHE $\gamma$-ray flux above 300\,GeV is shown in Fig.~\ref{fig4}. No significant variability between nights is found. The $\chi^2/\mathrm{d.o.f.}$ of the fit to a constant is 17.2/12.

\section{Multi-wavelength observations and results}

\rgb\ was observed in November 2007 by the Nan\c{c}ay Radio Telescope, the optical telescope ATOM located in the H.E.S.S. site, and the X-ray telescopes {\it RXTE}/PCA and {\it Swift}/XRT. This multi-wavelength campaign was triggered by the H.E.S.S. detection of \rgb\ thanks to target of opportunity agreements. We summarize here the results from these instruments.

\begin{itemize}
\item \textit{Swift}/XRT\\
  Observations between November 13, and 15, 2007. No variability was found. The averaged spectrum is described well by a broken powerlaw ($\Gamma_1 = 1.93 \pm 0.20$, $\Gamma_2 = 2.82 \pm 0.13$, $E_\mathrm{break} = 1.29 \pm 0.12$\,keV), with a flux of $F_{2-10\,\mathrm{keV}} \sim 2.7 \times 10^{-12}$\,erg\,cm$^{-2}$\,s$^{-1}$.
\item \textit{RXTE}/PCA\\
  Observations between November 13, and 15, 2007. No variability was found. The averaged spectrum is described by a powerlaw ($\Gamma = 2.72 \pm 0.08$). The resulting flux $F_{2-10\,\mathrm{keV}} \sim 6.8 \times 10^{-12}$\,erg\,cm$^{-2}$\,s$^{-1}$ exceeds the one obtained {\em simultaneously} with {\it Swift}. We attribute this to contamination by the nearby galaxy cluster Abell\,267 (see \cite{Kaufmann_Wagner} in these proceedings for more details on this issue).
\item Optical observations (ATOM)\\
  Observations between November 10, and 20, 2007. No variability was found. The host-galaxy-subtracted, core flux in the R-band was found to be $0.62 \pm 0.08$\,mJy.
\item Nan\c{c}ay Radio Telescope\\
  Observations on November 12, and 14, 2007. No variability was found. The flux at 2685\,MHz was found to be $56 \pm 6$\,mJy.
\end{itemize}

\section{SSC modelling}

\begin{figure}[h!]
  \centering
  \includegraphics[angle=-90,width=0.6\textwidth]{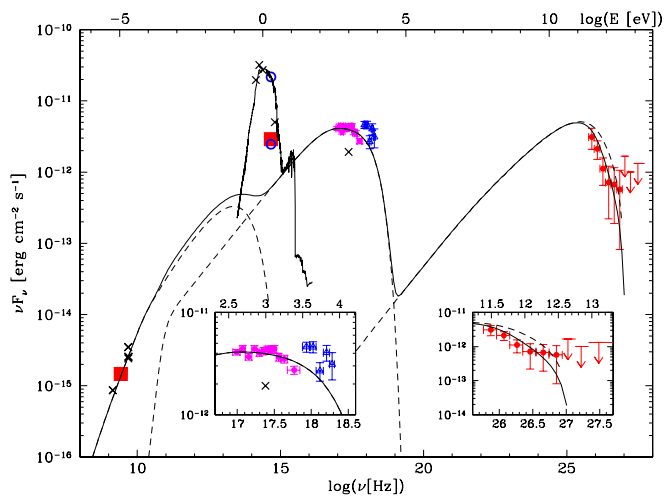}
  \caption{Broadband spectral energy distribution of \rgb. The contemporaneous data are shown in color. The solid lines show a 3 components model applied for this contemporaneous multi-wavelength campaign. Shown are the H.E.S.S. spectrum ({\it red filled circles and upper limits\/}), and contemporaneous {\it RXTE} ({\it blue open triangles\/}), {\it Swift}/XRT (corrected for Galactic absorption, {\it magenta filled circles}), optical host galaxy-subtracted (ATOM) and radio (Nan\c{c}ay) observations ({\it large red filled squares\/}). The black crosses are archival data. The contribution of the dominating host galaxy is shown in the optical band. The dashed line above the solid line at VHE shows the source spectrum after correcting for EBL absorption. The left- and right-hand side inlays detail portions of the observed X-ray and VHE spectrum, respectively.}
  \label{fig5}
\end{figure}

The simultaneous multi-wavelength (NRT, ATOM, {\it Swift}, {\it RXTE}, and H.E.S.S.) campaign allowed one to derive the SED of \rgb\ for the first time (see Fig.~\ref{fig5}), and to clearly confirm its HBL nature at the time of the H.E.S.S. observations.

Figure~\ref{fig5} also shows in solid lines an interpretation of the broadband emission of \rgb\ with 3 components:

\begin{itemize}
\item A simple non-thermal synchrotron model at low frequencies (radio) to describe the extended part of the jet.
\item The contribution of the dominating host galaxy in the optical, using a template of elliptic galaxy inferred from the code PEGASE \cite{1997A&A...326..950F}.
\item A leptonic non-thermal Synchrotron Self-Compton model describing the soft X-ray and VHE parts of the SED \cite{2001A&A...367..809K}.
\end{itemize}

\section{Conclusion}

The BL\,Lac \rgb\ was detected in VHE at energies $> 300$\,GeV with H.E.S.S. The contemporaneous multi-wavelength observations allow the SED of \rgb\ to be derived for the first time, clearly confirming its high-frequency-peaked nature at the time of H.E.S.S. observations.

\acknowledgments

The support of the Namibian authorities and of the University of Namibia in facilitating the construction and operation of H.E.S.S. is gratefully acknowledged, as is the support by the German Ministry for Education and Research (BMBF), the Max Planck Society, the French Ministry for Research, the CNRS-IN2P3 and the Astroparticle Interdisciplinary Programme of the CNRS, the U.K. Science and Technology Facilities Council (STFC), the IPNP of the Charles University, the Polish Ministry of Science and Higher Education, the South African Department of Science and Technology and National Research Foundation, and by the University of Namibia. We appreciate the excellent work of the technical support staff in Berlin, Durham, Hamburg, Heidelberg, Palaiseau, Paris, Saclay, and in Namibia in the construction and operation of the equipment.

This research made use of the SIMBAD database, operated at CDS, Strasbourg, France. This research also made use of the NASA/IPAC Extragalactic Database (NED). The authors thank the {\it RXTE} team for their prompt response to our ToO request and the professional interactions that followed. The authors acknowledge the use of the publicly available {\it Swift} data, as well as the public HEASARC software packages. This work uses data obtained at the Nan\c{c}ay Radio Telescope. The authors also thank Dr.~Mira V\'eron-Cetty for fruitful discussions.

\end{document}